\documentclass[
preprint,
sort&compress,
]{jasatex}

\usepackage{graphicx}
\usepackage{dcolumn}
\usepackage{amsmath,amsfonts,amssymb}
\usepackage{bm}
\begin{document}

\title[Backed Porous layer with inclusions]{Using simple shape three-dimensional inclusions to enhance porous layer absorption}

\author{J.-P. Groby}
\email{Jean-Philippe.Groby@univ-lemans.fr}
\affiliation{Laboratoire d'Acoustique de l'Universit\'e du Maine, UMR6613 CNRS/Univ. du Maine, Avenue Olivier Messiaen, F-72085 Le Mans Cedex 9, France.}
\author{B. Nennig}
\affiliation{Laboratoire d'Ing\'enierie des Syst\'emes M\'ecaniques et des Mat\'eriaux, (LISMMA EA2336), SUPMECA, 3 Rue Fernand Hainaut, 93407 Saint-Ouen Cedex, France.}
\author{C. Lagarrigue}
\affiliation{Laboratoire d'Acoustique de l'Universit\'e du Maine, UMR6613 CNRS/Univ. du Maine, Avenue Olivier Messiaen, F-72085 Le Mans Cedex 9, France.}
\author{B. Brouard}
\affiliation{Laboratoire d'Acoustique de l'Universit\'e du Maine, UMR6613 CNRS/Univ. du Maine, Avenue Olivier Messiaen, F-72085 Le Mans Cedex 9, France.}
\author{O. Dazel}
\affiliation{Laboratoire d'Acoustique de l'Universit\'e du Maine, UMR6613 CNRS/Univ. du Maine, Avenue Olivier Messiaen, F-72085 Le Mans Cedex 9, France.}
\author{V. Tournat}
\affiliation{Laboratoire d'Acoustique de l'Universit\'e du Maine, UMR6613 CNRS/Univ. du Maine, Avenue Olivier Messiaen, F-72085 Le Mans Cedex 9, France.}

\date{\today}

\begin{abstract}
The absorption properties of a metaporous material made of non-resonant simple shape three-dimensional inclusions (cube, cylinder, sphere, cone and torus) embedded in a rigidly backed rigid frame porous material is studied. A nearly total absorption can be obtained for a frequency lower than the quarter-wavelength resonance frequency due to the excitation of a trapped mode. To be correctly excited, this mode requires a filling fraction larger in the three-dimensions than in the two-dimensions for purely convex (cube, cylinder, sphere, and cone) shapes. At low frequencies, a cube is found to be the best purely convex inclusion shape to embed in a cubic unit cell, while the embedment of a sphere or a cone cannot lead to an optimal absorption for some porous materials. At fixed position of purely convex shape inclusion barycentre, the absorption coefficient only depends on and filling fraction and does not depend on the shape below the Bragg frequency arising from the interaction between the inclusion and its image with respect to the rigid backing. The influence of the angle of incidence is also shown. The results, in particular the excitation of the trapped mode, are validated experimentally in case of cubic inclusions.
\end{abstract}

\maketitle
\section{Introduction}\label{section1} 
Acoustic porous materials are widely used in noise control applications because of their good sound absorbing properties in the middle and high frequency range. Nevertheless, porous materials suffer from a lack of absorption at low frequencies, when compared to their absorption capabilities at higher frequencies. The usual way to solve this problem is by multi-layering, while trying to keep the thickness of the treatment relatively small compared to the incident wavelength that has to be absorbed. The purpose of the present article is to investigate an alternative to multi-layering by embedding non-resonant simple shape three-dimensional inclusions in a rigidly backed porous sheet, thus creating a diffraction grating and therefore extending to 3-dimensional configurations previous studies\cite{grobyjasa2011,nennigjasa2012} already conducted in 2-dimensional ones.

These last decades, several ways to avoid the problem of the porous material low frequency absorption have been proposed, mainly by combining resonant phenomena with the traditional viscous and thermal losses. Whatever the frequency, the idea behind is to excite modes of the structure that will trap the energy inside it and therefore enhance the absorption of the whole structure. The material properties were modeled either through homogenization procedures or by accounting for the whole wave phenomena: double porosity materials, whose properties are due to the microporous material resonance between the macro-pores\cite{olnyjasa2003}, or porous materials with small radius (compared to the wavelength) cylindrical inclusions embedded in\cite{tournatpre2004}, have been analyzed in the long wavelength limit, while metaporous materials\cite{grobyjasa2011,nennigjasa2012,clementnantes} have been studied either semi-analytically or numerically in the entire frequency range of audible sound. In particular, the effect of the periodic embedment of both non-resonant and resonant inclusions in a porous layer on the absorption properties were studied in two-dimensions when the porous layer either is backed by a rigid backing\cite{grobyjasa2011,nennigjasa2012,clementnantes}, possibly incorporating cavities\cite{grobyjasa2011b}, or radiates in a semi-infinite half-space\cite{grobywrcm2008}. Different inclusion shapes were studied\cite{grobyjasa2011,nennigjasa2012,clementnantes,hyunjasa2012} showing similar results at low frequencies. The increased absorption was explained by the excitation of local mode of the inclusion or of the cavities of the rigid backing, by the excitation of a trapped mode (TM) that traps the energy between the inclusion and the rigid backing, and by the excitation of the modified mode of the layer by coupling the layer mode with the Bloch waves originated by the added periodic heterogeneities. The effects of the inplane periodicity, whose main effect is the different excitation of modified mode of backed layer (MMBL), discussed in detail in \cite{grobyjasa2013} in case of parallelepipedic irregularities of the rigid backing, will not be investigated here.

In this article, the influence of the periodic embedment of three-dimensional elementary shape inclusions in a porous layer rigidly backed is studied by use of an in-house Finite Element (FE) code. These simple shape three-dimensional inclusions cover a wide range of topological characteristics. For instance, cube, cylindre, sphere and cone have purely convex geometry, while the torus is not convex and presents some concave faces. Furthermore, the cone does not possess geometric symmetry with respect to its barycenter contrary to the other. The absorption coefficient of the whole structure for different shapes and orientations of the inclusions is calculated. Only square lattice are considered, i.e., the periodicities are identical in both directions of the plane.

The present paper is organized as follows. The problem is described in section II. The FE method is then validated numerically and experimentally in section III. In section IV, various numerical examples with different inclusion shapes are discussed.

\section{Formulation of the problem}\label{section2}

\subsection{Description of the configuration}\label{section2ss1}
\begin{figure}
\includegraphics[width=.49\textwidth]{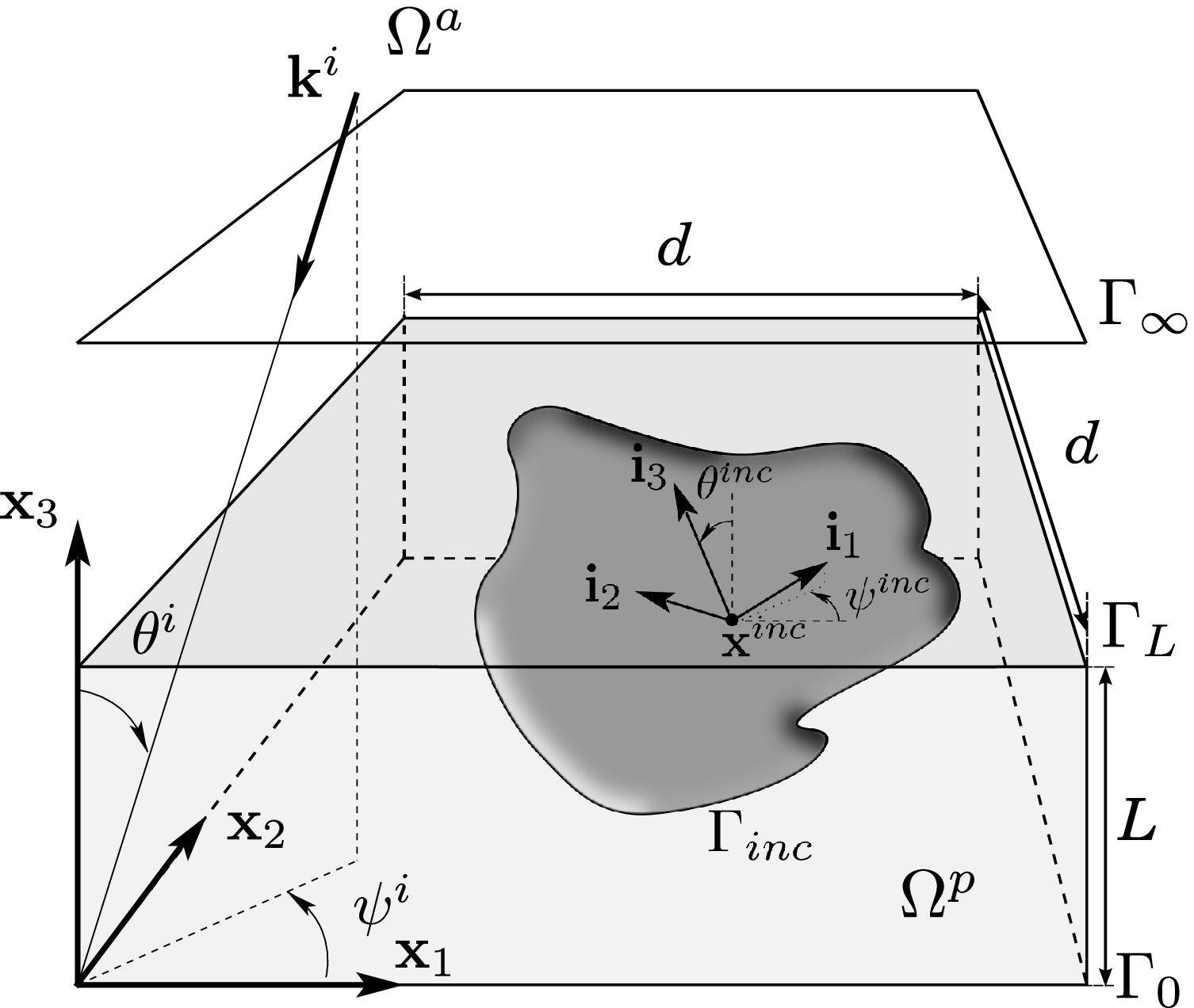}
\caption{Example of a $\mathbf{d}$-periodic fluid-like porous sheet backed by a rigid wall with a periodic inclusion embedded in.} 
\label{section2ss1fig1}
\end{figure}

A parallelepipedic unit cell of the 3D scattering problem is shown in Fig. \ref{section2ss1fig1}. Before the addition of the inclusions, the layer is a rigid frame porous material saturated by air (e.g., a foam) which is modeled as a macroscopically homogeneous equivalent fluid $M^{p}$ using the Johnson-Champoux-Allard model\cite{johnson,allardchampoux}. The upper and lower flat and mutually parallel boundaries of the layer, whose $x_{3}$ coordinates are $L$ and $0$, are designated by $\Gamma_{L}$ and $\Gamma_{0}$ respectively. The upper semi-infinite material $M^{a}$, i.e., the ambient fluid that occupies $\Omega^{a}$, and $M^{p}$ are in a firm contact at the boundary $\Gamma_{L}$, i.e., the pressure and normal velocity are continuous across $\Gamma_{L}$. A Neumann type boundary condition is applied on $\Gamma_0$, i.e. the normal velocity vanishes on $\Gamma_0$.

\begin{figure}
\includegraphics[width=12.0cm]{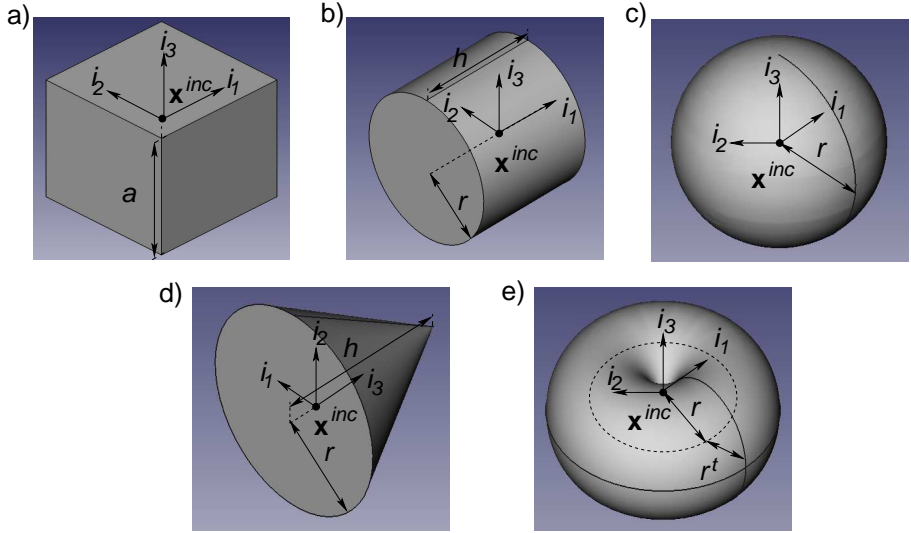}
\caption{Sketch of the different simple shape inclusions considered (of barycenter $\mathbf{x}^{inc}$): cubic inclusions (a), cylindrical inclusions (b), spherical inclusions (c), conic inclusion (d) and toric inclusion (e)} 
\label{section2ss1fig2}
\end{figure}

Inclusions, with a common spatial periodicity $\mathbf{d}=<d_1,d_2,0>$, are embedded in the porous layer and create a two-dimensional diffraction grating in the plan $x_1-x_2$. The periodicities $d_1$ and $d_2$ along the $x_1$ and $x_2$ directions are identical, $d_1=d_2=d$. In the following, five different infinitely-rigid simple-shape inclusions, depicted in Fig. \ref{section2ss1fig2}, are considered : a cubic inclusion of edge $a$, a cylindrical inclusion of radius $r$ and length $h$, a spherical inclusion of radius $r$, a conic inclusion of radius $r$ and height $h$, and toric inclusions of neutral axis radius $r$ and tore radius $r^t$. A Cartesian coordinate system, with the three unit vectors $\mathbf{i}_j$, $j=1,2,3$, is attached to each inclusion barycenter. The position and orientation of the inclusion are refered to by its barycenter $\mathbf{x}^{inc}$, its azimuth $\psi^{inc}$, and its elevation $\theta^{inc}=(\mathbf{i}_3,\mathbf{x}_3)$.
 
The incident wave propagates in $\Omega^{a}$ and is expressed by $\displaystyle p^{i}(\mathbf{x})=A^{i}e^{\textrm{i}(k_{1}^{i}x_{1}+k_{2}^{i}x_{2}-k_{3}^{ai}(x_{3}-L))}$, wherein $k_1^i=-k^{a}\sin \theta^i \cos \psi^i$, $k_2^i=-k^{a}\sin \theta^i \sin \psi^i$,  $k_3^{ai}=k^{a}\cos \theta^i$ and $A^{i}=A^{i}(\omega)$ is the signal spectrum. The azimuth of the incident wavevector is $\psi^{i}$ and its elevation $\theta^{i}$. 

In each domain $\Omega^{\alpha}$ ($\alpha=a,p$), the pressure field fulfills the Helmholtz equation
\begin{equation}\label{eq:Helmholtz}
\nabla \cdot \left( \frac{1}{\rho^{\alpha}} \nabla p^{\alpha} \right) + \frac{(k^{\alpha})^2}{\rho^{\alpha}} p^{\alpha} = 0,
\end{equation}
with the density $\rho^{\alpha}$ and the wavenumber $k^{\alpha}=\omega/c^{\alpha}$, defined as the ratio between the angular frequency $\omega$ and the sound speed $c^{\alpha}$.

As the problem is periodic and the excitation is due to a plane wave, each field ($X$) satisfies the Floquet-Bloch relation
\begin{equation} \label{eq:Floquet}
X(\mathbf{x}+\mathbf{d}) = X(\mathbf{x}) e^{\textrm{i}\mathbf{k}^{i}_\bot \cdot \mathbf{d}  },
\end{equation}
where $\mathbf{k}^{i}_\bot = <k_1^{i},\, k_2^{i},\, 0>$ is the in-plane component of the incident wavenumber. Consequently, it suffices to examine the field in the elementary cell of the material to get the fields, via the Floquet relation, in the other cells. The periodic wave equation is solved with a FE method. This FE method as well as the absorption coefficient calculation method are described in the appendix.

\subsection{Material modeling}\label{section2ss2}
The rigid frame porous material is modeled using the Johnson-Champoux-Allard model. The compressibility and density, linked to the sound speed through $c^p=\sqrt{1/\left(K^p\rho^p\right)}$ are\cite{johnson,allardchampoux}
\begin{equation}
\begin{array}{ll}
\displaystyle \frac{1}{K^p}=&\displaystyle\frac{\gamma P_0}{\displaystyle \phi\left(\gamma-\left(\gamma-1 \right)\left(1+\textrm{i}\frac{\omega_c'}{\mbox{Pr }\omega}G(\mbox{Pr }\omega) \right)^{-1} \right)}~,\\[24pt]
\displaystyle \rho^p=&\displaystyle\frac{\rho^a \alpha_{\infty}}{\phi}\left(1+\textrm{i}\frac{\omega_c}{\omega}F(\omega) \right)~,
\end{array}\label{section2ss2e1}
\end{equation}
wherein $\displaystyle \omega_c=\sigma \phi/\rho^a \alpha_{\infty}$ is the Biot frequency, $\displaystyle \omega_c'=\sigma' \phi/\rho^a \alpha_{\infty}$ is the adiabatic/isothermal crossover frequency, $\gamma$ the specific heat ratio, $P_0$ the atmospheric pressure, $\mbox{Pr}$ the Prandtl number, $\rho^a$ the density of the fluid in the (interconnected) pores, $\phi$ the porosity, $\alpha_\infty$ the tortuosity, $\sigma$ the flow resistivity, and $\sigma'$ the thermal resistivity. The correction functions $G(\mbox{Pr }\omega)$ \cite{allardchampoux} and $F(\omega)$ \cite{johnson} are given by
\begin{equation}
\begin{array}{ll}
\displaystyle G(\mbox{Pr }\omega)=&\displaystyle \sqrt{1-\textrm{i} \eta \rho^a \mbox{Pr }\omega \left(\frac{ 2\alpha_{\infty}}{\sigma' \phi \Lambda'}\right)^2} ~,\\[8pt]
\displaystyle F(\omega)=&\displaystyle \sqrt{1-\textrm{i} \eta \rho^a \omega \left(\frac{ 2\alpha_{\infty}}{\sigma \phi \Lambda}\right)^2    }~,
\end{array}
\label{section2ss2e2}
\end{equation}
where $\eta$ is the viscosity of the fluid, $\Lambda'$ the thermal characteristic length, and $\Lambda$ the viscous characteristic length. The thermal resistivity is related to the thermal characteristic length\cite{allardchampoux} through $\sigma'=8 \alpha_\infty \eta/\phi \Lambda'^2$.
\section{Numerical and experimental validation}
A large tortuosity ($\alpha_\infty=1.42$) $20\mbox{ mm}$ thick foam (Fireflex) sheet $S1$ and a medium resistivity ($\sigma=11500 \textrm{ N.s.m}^{-4}$) $22\mbox{ mm}$ thick foam (Melamine) sheet $S2$ are used thorough the article. The parameters of these two porous materials are reported in Table \ref{table1}. These parameters have been evaluated using the traditional methods (Flowmeter for the resistivity and ultrasonic methods for the $4$ other parameters, together with a cross-validation by impedance tube measurement) described in \cite{Allard}.
\begin{table}
\begin{ruledtabular} 
\begin{center}
\begin{tabular}{cccccc}
& $\phi$ & $\alpha_\infty$ & $\Lambda$ ($\mu$m) & $\Lambda'$ ($\mu$m) & $\sigma$ (N.s.m$^{-4}$) \\
\hline
S1 & 0.95 & 1.42 & 180 & 360 & 8900 \\
S2 & 0.99 & 1.02 & 120 & 240 & 11500 \\
\end{tabular}
\end{center}
\caption{Acoustical parameters of the porous material constituting the sheet of thickness $L$.}
\label{table1}
\end{ruledtabular}
\end{table}

\begin{figure}			
\includegraphics[width=8.0cm]{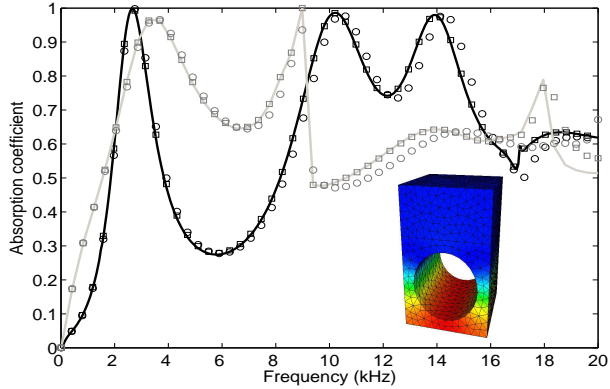}
\caption{Absorption coefficient for an infinitely long cylinder inclusion, when the layer is occupied by the foam S1, see Table \ref{table1}, excited at normal incidence (black) or with $\theta^{i}=\pi/3$ (gray): reference results from Ref.~~\onlinecite{grobyjasa2011}, i.e., 2D results (solid line), linear elements (o) and quadratic elements ($\square$). The inset shows the snapshot of the pressure field magnitude at 2674 Hz.}
\label{fig:ValidCircle}
\end{figure}

Extruded 2D configuration has first been used to validate the proposed 3D FE method by comparison with 2D results. This 2D configuration has been extensively validated with the multipole method\cite{grobyjasa2011}, with a modal approach\cite{nennigjasa2012} and with FE method\cite{grobyjasa2011}. This 2D FE method was based on a slightly different approach. This configuration consists in a $2\textrm{ cm}$ thick foam S1 with rigid circular inclusion of radius $r=7.5\textrm{ mm}$ embedded in with a spatial periodicity $d=2\textrm{ cm}$. The comparision of the absorption coefficient calculated with the present FE method and the multipole method is presented in Fig.\ref{fig:ValidCircle}, showing a good agreement for both oblique an normal incidences. Around 10 linear elements per edges ($2\textrm{ cm}$) on the elementary cell leads to less than 1\% of error on the absorption coefficient below $10\textrm{ kHz}$ and around 5\% above. With the same mesh, quadratic elements yield to less than 1 \% of error on the absorption curves thorough $20\textrm{ kHz}$. Quadratic elements are used through this paper when calculations are run through $20\textrm{ kHz}$, while linear elements are used when calculations are run for frequencies lower than $10\textrm{ kHz}$. These results are in line with standard FE rules of thumbs, i.e., 10 linear elements per wavelength for around 1\% of error.

\begin{figure}
\includegraphics[width=8.0cm]{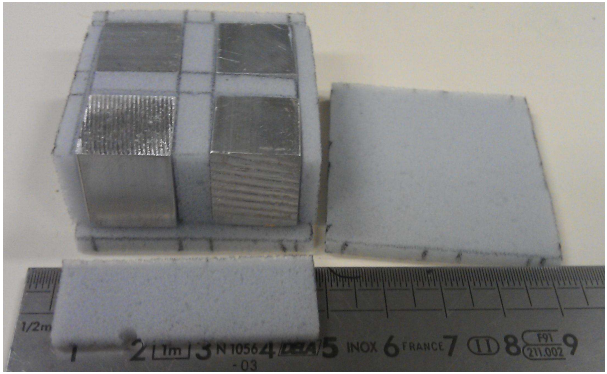}
\caption{Picture of the characterized sample, configuration C2: a $2.2\textrm{ cm}$ thick Melamine foam with four $15\textrm{ mm}$ edge cube embedded in.}
\label{section3ss2fig1}
\end{figure}

The proposed method has also been validated by comparison with experimental results at normal incidence. The tested sample is composed of a Melamine foam (S2, Table \ref{table1}) as the porous matrix and four aluminum cubes of $15\textrm{ mm}$ length edges as shown Fig.\ref{section3ss2fig1}. The sample also contains $4$ unit cells. The initial $22\textrm{ mm}$-thick melamine foam was sliced and the material volumes that are then occupied by the inclusions were removed. The different elements are then gathered together with thin glue layers. The sample is placed at the end of an impedance tube, with a square cross-section and a side length of $42\textrm{ mm}$, against a copper plug that closes the tube and acts as a rigid boundary. The tube cut-off frequency is $4200\textrm{ Hz}$. By assuming that only plane waves propagate below the cut-off frequency, the infinitely rigid boundary conditions of the tube act like perfect mirrors and create a periodicity pattern in the $x_1$ and $x_2$ directions with a periodicity of $21\textrm{ cm}$, because $4$ inclusions are embedded in the sample. This technique was previously used in\cite{grobyjasa2013,clementnantes} and allows to determine experimentally the absorption coefficient of a quasi-infinite inplane periodic structure just with one or a correctly arranged small number of unit cells. 

\begin{figure}
\includegraphics[width=8.0cm]{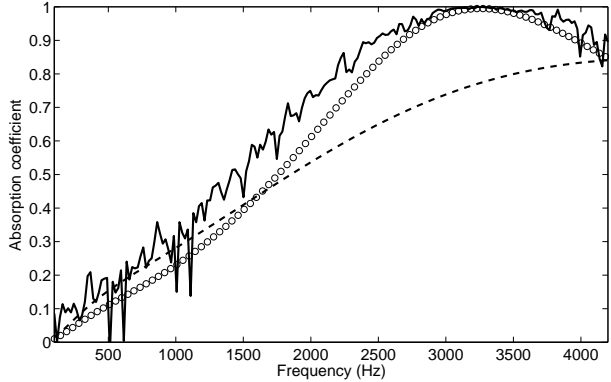}
\caption{Absorption coefficient (linear elements) of a $15\textrm{ mm}$ edge cube, when the layer is occupied by the foam S2, excited at normal incidence: experimental results (solid line), numerical results (o), and homogneous layer (dashed line).}
\label{section3ss2fig2}
\end{figure}

Figure \ref{section3ss2fig2} shows a comparison between the absorption coefficient of this sample measured experimentally and calculated with the present FE method. Both absorption coefficients are in good agreement. The small differences can be attributed to the glue layers and possible thin air  layers inside the sample. The absorption coefficient of the corresponding homogeneous layer is also depicted in Figure \ref{section3ss2fig2}. Measurement of the initial $2.2\textrm{ cm}$-thick foam, without inclusions were also performed showing a perfect agreement with the model. These measurements are not shown here for clarity of the figure. The experiments show an increase of the absorption coefficient (almost total) at low frequency due to the excitation of the trapped mode.
Other experiments, not shown, were performed in the case of a vertical cylinder embedded in, also showing a good agreement with the calculations, therefore validating the FE calculations experimentally.
\section{Numerical results}\label{section3}

Numerical simulations have been performed for various geometric parameters, various shape inclusions and within the frequency range of audible sound, particularly at low frequencies. One of the main constraints in designing acoustically absorbing materials is the size and weight of the configuration. In this sense, a low frequency improvement implies good absorption for wavelength larger than the thickness of the structure. 

The dimensions of the main studied configurations are listed in Table \ref{table2}.
\begin{table*} 
\begin{ruledtabular} 
\begin{center}
\begin{tabular}{ccccccc}
Configuration & $d$ (mm) & $L$ (mm) & Inclusion type &  Inclusion dimensions (mm) & $\theta^{inc}$, $\psi^{inc}$ & ($x_1^{inc}$, $x_2^{inc}$, $x_3^{inc}$)\\
\hline
C1 & 20 & 20 & Cube & a=16 & 0, 0 & (10,~10,~10) \\
C2 & 20 & 20 & Cube & a=15 & 0, 0 & (10,~10,~10) \\
C3 & 20 & 20 & Cylinder & $h=15$, $r=8.5$ & & (10~,10~,10) \\
C4 & 20 & 20 & Sphere & $r=9.3$ & 0, 0 & (10,~10,~10) \\
C5 & 20 & 20 & Cone & $r=8.5$, $h=15$ &  &  \\
C6 & 20 & 20 & Torus & $r=5$, $r^t=4.75$ &  & (10~,10,~10) \\
\end{tabular}
\end{center}
\caption{Dimensions of the main studied configurations.}
\label{table2}
\end{ruledtabular}
\end{table*}
\subsection{Cubic inclusions}\label{section3ss1}

\begin{figure}
\includegraphics[width=8.0cm]{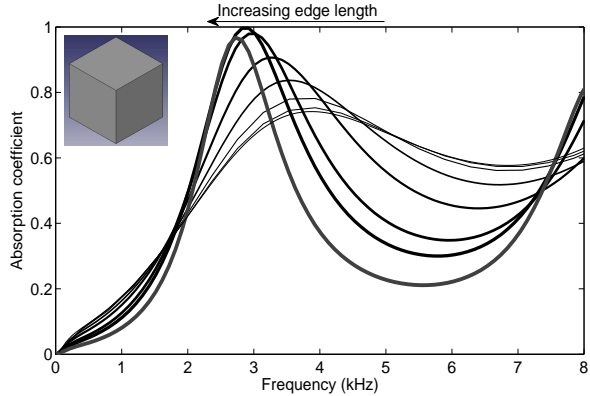}
\caption{Absorption coefficient (linear elements) of a cubic inclusion of edge length $a$ centered in a cubic unit cell $(d_1,d_2,L)=(2\textrm{ cm},2\textrm{ cm},2\textrm{ cm})$ occupied by the foam S1 when excited at normal incidence: from the thinest to the thickest curves $a=0$, $a=5\textrm{ mm}$, $a=7.5\textrm{ mm}$, $a=10\textrm{ mm}$, $a=12.5\textrm{ mm}$, $a=15\textrm{ mm}$, $a=16\textrm{ mm}$, and $a=17.5\textrm{ mm}$.} 
\label{section3ss1fig1}
\end{figure}

First, a $a$-edge cubic inclusion is considered in a cubic unit cell $(d,d,L)=(2\textrm{ cm},2\textrm{ cm},2\textrm{ cm})$. The cube is centered in the unit cell, i.e. $\mathbf{x}^{inc}=(d/2,d/2,L/2)$. The inclusion is oriented such that $(\theta^{inc},\psi^{inc})=(0,0)$, i.e., the faces of the cube are parallel to those of the unit cell. Figure \ref{section3ss1fig1} depicts the evolution of the absorption coefficient at normal incidence for various edge lengths $a$ from $0\textrm{ mm}$ to $17.5\textrm{ mm}$ leading to different filling fractions $f\!f$ from $0$ to $\approx 0.67$.

Similarly to the analysis carried out in the two-dimensional case\cite{grobyjasa2011,nennigjasa2012}, a trapped mode (TM) is excited by the presence of the inclusion. The TM excitation frequency $\nu^{t}$ becomes lower when the filling fraction increases for fixed position of the barycenter. This frequency is always lower than the quarter wavelength resonance one, when the barycenter is higher or equal to half of the layer thickness $L$. The absorption coefficient possesses a maximum as a function of the edge of the cube. The absorption is nearly total for $a=16\textrm{ mm}$ edge cubic inclusion (configuration C1), which corresponds to a filling fraction of $f\!f\approx 0.51$. When compared to the results obtained in the 2D case\cite{grobyjasa2011} for a $r=7.5\textrm{ mm}$ infinitely long cylinder centered in the same matrix material with identical geometry (Figure \ref{fig:ValidCircle}), the required filling fraction for the absorption peak to be total is larger in the 3D case, i.e. $f\!f\approx 0.51$, than in the 2D case, $f\!f\approx 0.44$. In the same way, when the nearly total absorption peak is reached, $\nu^{t}$ is higher in the 3D case ($\nu^{t} = 2860\textrm{ Hz}$) than in the 2D case ($\nu^{t} = 2680\textrm{ Hz}$). 

\begin{figure}
\includegraphics[width=6.0cm]{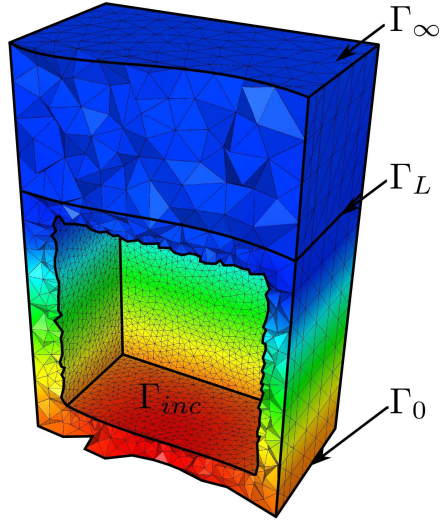}
\caption{Snapshot of the pressure field magnitude (linear elements) along a cross-sectional ($x_1-x_3$ plan) plan view at $x_2=d/2$ in the configuration C1, when the layer is occupied by the foam S1, excited at normal incidence at $\nu=2860\textrm{ Hz}$.} 
\label{section3ss1fig2}
\end{figure}

Figure \ref{section3ss1fig2} depicts a cross-sectional view ($x_1-x_3$ plane) of the pressure field magnitude inside the unit cell at $x_2=d/2$ of the configuration C1 at $\nu^{t} = 2860\textrm{ Hz}$, showing that the wave is trapped between the inclusion and the rigid backing. Similarly to the 2D case, at fixed edge $a$, $\nu^{t}$ becomes smaller when the distance between the inclusion and the rigid backing is larger, i.e., when $x_3^{inc}$ increases. Therefore, this increased absorption could be explained by the first Fabry-Perot interference between the inclusions and its image with respect to the rigid backing, as can be shown in the transmission case. Nevertheless, the Fabry-Perot interference appears when the vertical distance between two adjacent inclusions is equal to the quarter of the projection on the vertical axis of the wavevector, which is impossible for two reasons: 1. because a quarter-wavelength is not possible with the rigid backing 2. because $\nu^t$ would be identical for each cube edges and equal to the quarter-wavelength resonance frequency for a centered inclusion, if true.  

\begin{figure}
\includegraphics[width=8.0cm]{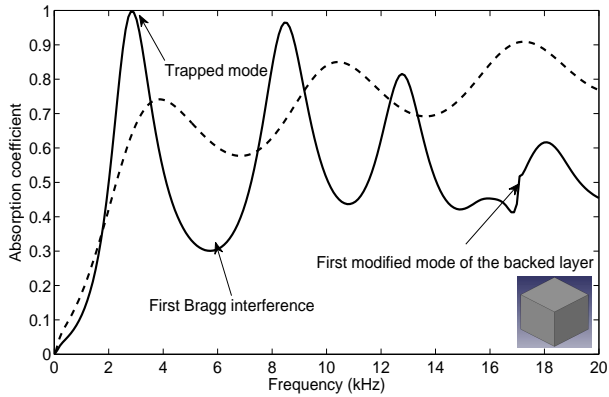}
\caption{Absorption coefficient (quadratic elements) of the configuration C1, when the layer is occupied by the foam S1, excited at normal incidence.  The absorption coefficient of the homogeneous layer is depicted by the dashed line.}
\label{section3ss1fig3}
\end{figure}

Once the optimal edge size is determined to have a nearly total absorption peak at $\nu^{t}$, the absorption coefficient is calculated for the total frequency range of audible sound in Fig. \ref{section3ss1fig3}. The first Bragg interference, which corresponds to the maximum of reflected energy leads to a minimum of absorption around $6000\textrm{ Hz}$. This corresponds to constructive interferences between the scattered waves by the inclusion and its image with respect to the rigid backing. This minimum appears when $2 x_{3}^{inc}$ is equal to half of the wavelength in case of normal incidence. 

The modified mode of the backed layer (MMBL), which traps the energy inside the porous plate and corresponds to an evanescent wave in the upper half plate and a propagative wave inside the porous plate is excited at $\nu^{MMBL}\approx 17000\textrm{ Hz}$. This corresponds to the intersection of the longitudinal mode of the porous plate, which cannot be excited by a plane incident wave without heterogeneity, with the first Bloch wave, as explained in\cite{grobyjasa2010,grobyjasa2011,grobyjasa2013}. This mode is excited at relatively high frequency because the periodicity is here relatively small ($d=2\textrm{ cm}$). The enhanced absorption due to MMBL was extensively explained in case of parallelepiped irregularities of the rigid backing in\cite{grobyjasa2013} and used in\cite{miguelny}.

The absorption coefficients calculated in the case of a centered $a=12.5\textrm{ mm}$ ($f\!f\approx 0.24$) edge cube with upper and lower interfaces parallel to $\Gamma_L$ and $\Gamma_0$ rotated around $\mathbf{i}_3$ are almost identical below the first MMBL. In particular, the absorption coefficients calculated for $\psi^{inc}=0$ and $\psi^{inc}=\pi/4$ ($\theta^{inc}=0$) are identical, while it is well known, \cite{pichardjasa2012,goffauxprb2001}, that a 2D sonic crystal composed of square cross-section scatterers possesses full bandgap when $\psi^{inc}=\pi/4$ and only a bandgap at normal incidence ($\Gamma X$), when $\psi^{inc}=0$. Some differences can be noticed near grazing incidence, but are not significant in case of an acoustic excitation by an airborne plane wave impiging the structure from the upper half-space. The bandgap were shown and were clearly of interest because the excitation was performed in between the parallelepipedic scatterers in \cite{MiguelJAP}.

Parallelepiped scatterers were tested, exhibiting similar influence on the absorption. The advantage is that $\nu^t$ can be smaller because the parallellepiped can be placed further from the hard backing $\Gamma_0$, letting $x_3^{inc}$ being larger, but the filling fraction is then lower and a nearly total absorption peak is then difficult to reach at this frequency.
\subsection{Other simple shape inclusions at normal incidence}\label{section3ss3}

We first focus on the three purely convex simple inclusion shapes that possesse geometric symmetry with respect to their barycenter.

The absorption coefficients for different simple shape inclusions embedded in the same porous material with the same periodicity and at identical filling fraction are calculated and compared. For identical material layer and dimension of the unit cell, the large filling fraction required, $f\!f\approx 0.51$, to reach an almost total absorption coefficient at $\nu^t$ in case of cube would impose a $r\approx 9.9\textrm{ mm}$ radius sphere. Such a filling fraction is also impossible to realize in practice when spheres are embedded with a periodicity $d=2\textrm{ cm}$. This also means that in a cubic unit cell, a cubic inclusion seems to be the best choice in the sense that this inclusion shape enables large filling fraction when compared with other simple shape inclusions, like sphere or cylinder. While the embedment of spheres in a porous materials probably do not lead to an optimal absorption, i.e. a unit amplitude absorption, at low frequency, this solution is used in practice\cite{fuller2012}.

\begin{figure}
\includegraphics[width=8.0cm]{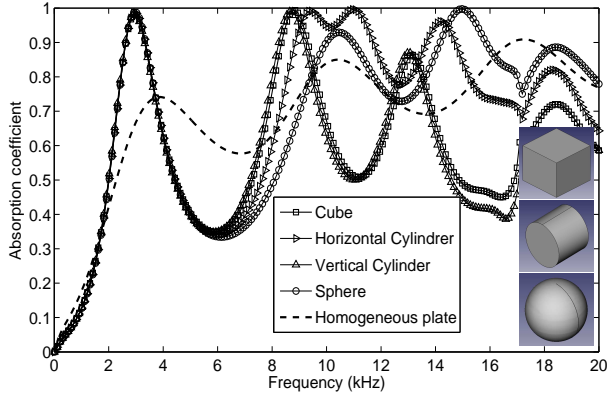}
\caption{Absorption coefficients (quadratic elements) of the configuration C2 ($\square$), C3 $\theta^{inc}=0$ ($\triangleright$) and $\theta^{inc}=\pi/2$ ($\triangle$), and C4 (o), when the layer is occupied by the foam S1, excited at normal incidence. The absorption coefficient of the homogeneous layer is depicted by the dashed line.}
\label{section3ss3fig1}
\end{figure}

The cube edge is also decreased to $a=15\textrm{ mm}$ in order for the filling fraction to be $f\!f\approx 0.42$. A comparison between the absorption coefficients calculated for centered inclusions in the unit cell of a $a=15\textrm{ mm}$ cube (configuration C2), $h=15\textrm{ mm}$ and $r=8.5\textrm{mm}$ cylinder with different orientations (configuration C3: $\theta^{inc}=0$ vertical cylinder and $\theta^{inc}=\pi/2$ horizontal cylinder), and a $r=9.3\textrm{ mm}$ sphere (configuration C4) is shown in figure \ref{section3ss3fig1}. Several observations can be made. First, for fixed properties of material layer, inclusion barycenter position periodicity and filling fraction, the absorption coefficient is identical for the different inclusion shapes below the first Bragg frequency. This means that the absorption coefficient is mainly driven by the filling fraction below this frequency. Second, whatever the inclusion shape, the MMBL is excited at the same frequency $\nu^{MMBL}$, because it only depends on the layer material properties and periodicity. Thirdly, absorption coefficients are almost identical between a vertical cylinder (configuration C3, $\theta^{inc}=0$) and a cube (configuration C2). This means that at higher frequency flat interfaces parallel with the boundaries $\Gamma_0$ and $\Gamma_L$ have more influence on the absorption coefficient than the lateral shape. Fourthly, absorption coefficients for flat interface inclusions parallel with $\Gamma_L$ and $\Gamma_0$ are completely different from those for non-flat interface inclusions. Fifthly, while cubic inclusion enable a larger filling fraction, spherical and horizontal cylindrical inclusions lead to larger absorption at higher frequency. The higher order Bragg interferences seem to be more excited than in case of cube and vertical cylinder. This is in accordance with the conclusion of \cite{wangjap2001} in which it is shown in 2D that square cross-section scatterer in square lattice provides a larger bandgap, than other scatterer shapes. The absorption is almost always larger than the one of the homogeneous layer between the Bragg frequency and the first MMBL frequency for spheres and horizontal cylinders. In average, the best absorption coefficient is obtained with an horizontal cylindrical inclusion.

Calculations were also perform for each inclusion shape when the filling fraction increases. For each inclusion shape, $\nu^t$ decreases and the absorption amplitude increases with increasing $f\!f$ at fixed barycenter position. For cylinders, it was possible to find a possible to manufacture configuration leading to a nearly total absorption peak, while it was not the case of a sphere. Once the optimum is reached, if the filling fraction $f\!f$ is still increased, both frequency $\nu^t$ and amplitude of the associated absorption peak decrease.

The last two shapes studied here present particularities: the cone does not present geometric symmetry with respect to its barycenter and the torus is not of convex shape.

\begin{figure}
\includegraphics[width=8.0cm]{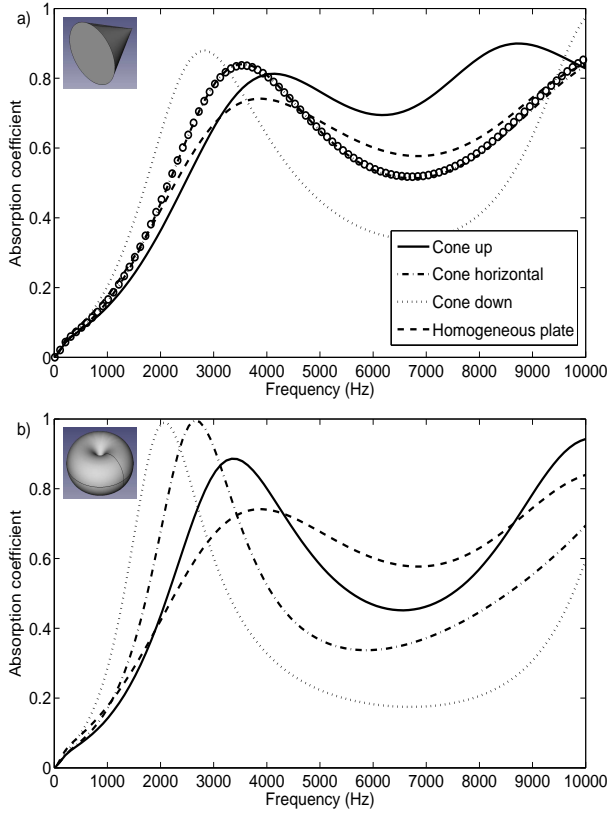}
\caption{Absorption coefficients of a) a $h=15\textrm{ mm}$ and $r=8.5\textrm{ mm}$ cone, configuration C5, in a $2\textrm{ cm}$ thick S1 layer, when the cone is up $\theta^{inc}=0$, cone horizontal $\theta^{inc}=-\pi/2$, and cone down $\theta^{inc}=\pi$ (the absorption coefficient of a $a=10\textrm{ mm}$ cube is also plotted (o)) and b) a $r=5\textrm{ mm}$ and $r^t=4.75\textrm{ mm}$ torus, configuration C6 in a $2\textrm{ cm}$ thick S1 layer, centered in the unit cell when $\theta^{inc}=0$ (dashed-dotted line) and $\theta^{inc}=\pi/2$ (solid line), and when $x_3^{inc}=3L/4$ for $\theta^{inc}=0$ (dotted line). The absorption coefficient of the homogeneous layer is depicted by the dashed line on a) and b).}
\label{section3ss3fig2}
\end{figure}

Concerning conic inclusions, besides an apparent dependance of $\nu^t$ on the orientation of the cone, all the phenomena are in accordance with the previously studied shapes. Figure \ref{section3ss3fig2}(a) depicts the absorption coefficient of a $h=15\textrm{ mm}$, $r=8.5\textrm{ mm}$ cone, configuration C5, embedded in a porous sheet S1, and geometrically centered in the unit cell for different orientations: cone up, $\mathbf{x}^{inc}=(10 \textrm{ mm},10 \textrm{ mm},6.25 \textrm{ mm})$, $\theta^{inc}=0$; cone horizontal, $\mathbf{x}^{inc}=(6.25 \textrm{ mm},10 \textrm{ mm},10 \textrm{ mm})$, $\theta^{inc}=-\pi/2$; and cone down, $\mathbf{x}^{inc}=(10 \textrm{ mm},10 \textrm{ mm},13.25 \textrm{ mm})$, $\theta^{inc}=\pi$. The corresponding filling fraction is $f\!f\approx0.15$. It is impossible to reach a sufficiently large filling fraction to obtain a nearly total absorption peak with the conic inclusion in this case. When the cone is horizontal, the $x_3$-coordinate of the barycenter is located in the middle height unit cell, i.e. $x_3^{inc}=L/2$. The configuration being periodic, it is possible to find a unit cell such that the barycenter is the center of the unit cell. In this case, the absorption peak associated with the excitation of the trapped mode is very close to the one of a $a=10\textrm{ mm}$ cube, $f\!f= 0.125$ (rigorously the edge of the cube should be $a=10.5\textrm{ mm}$ for $f\!f\approx0.15$). When the cone is oriented towards the air medium, i.e. cone up, the barycenter is lower than the middle of the unit cell, while when the cone is oriented towards the rigid backing, i.e. cone down, the barycenter is higher than the middle of the unit cell. This dependence is identical to the one already noticed in case of a cube, a sphere and a cylinder, i.e. $\nu^t$ decreases when $x_3^{inc}$ increases.

The toric shape is more interesting and more complex. The absorption coefficients of $r=5\textrm{ mm}$ and $r^t=4.75\textrm{ mm}$ torus (configuration C6) embedded in the same S1 porous sheet, centered in the unit cell, i.e., $\mathbf{x}^{inc}=(10\textrm{ mm},10\textrm{ mm},10\textrm{ mm})$, with two different orientations, $\theta^{inc}=0$ and $\theta^{inc}=\pi/2$ ($\psi^{inc}=0$), are shown in Figure \ref{section3ss3fig2}(b). For this shape, a nearly total absorption peak can be obtained at $\nu^t=2680\textrm{ Hz}$ and for a filling fraction $f\!f\approx 0.28$, i.e., both frequency and filling fraction are lower than those obtained in the case of purely convex shapes. This phenomenon is due to the non-convex shape which enables the volume below the inclusions to be larger in case of a torus than in case of purely convex shapes. This relatively low filling fraction associated with the toric shape also allows to reach a nearly total absorption peak at a very low frequency ($\nu^t=2100\textrm{ Hz}$) when $x_3^{inc}$ is increased at $x_3^{inc}=3L/4$, Fig. \ref{section3ss3fig2}(b). Nevertheless, this last shape should be considered as a particular case of a simple 3D shape, and can be considered at high frequency as a resonant one.

\subsection{Numerical results at oblique incidence}\label{section3ss4}

\begin{figure}
\includegraphics[width=8.0cm]{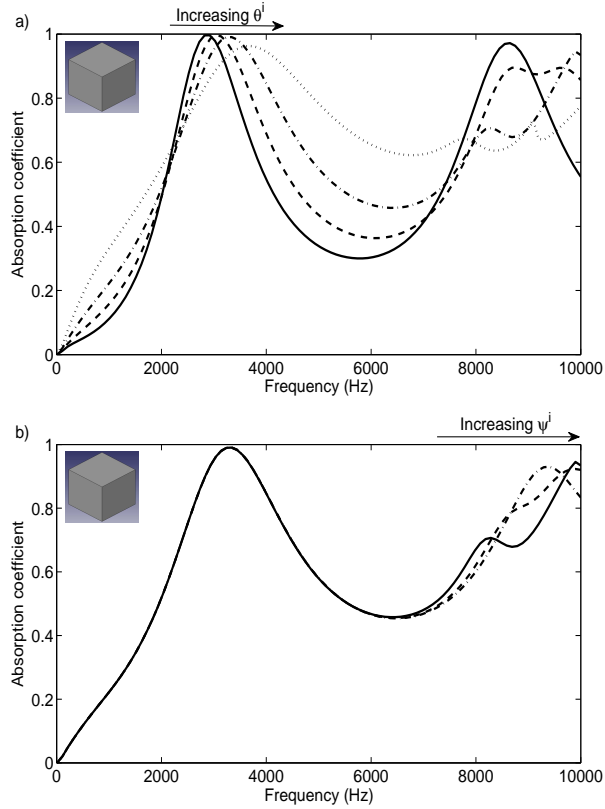}
\caption{Absorption coefficient of a layer with cubic inclusions of $a=16\textrm{ mm}$ edge length centered in a cubic unit cell $(d,L)=(2\textrm{ cm},2\textrm{ cm})$ occupied by the foam S1 when excited a) for $\psi^i=0$ and $\theta^i=0$, $\theta^i=\pi/6$, $\theta^i=\pi/4$, and $\theta^i=\pi/3$, and b) for $\theta^i=\pi/4$ and $\psi^i=0$, $\psi^i=\pi/6$, and $\psi^i=\pi/4$.}
\label{section2ss1fig3}
\end{figure}

The structure is obviously anisotropic, first because of the periodicity patterns itself, but also because of the inclusion shape. 
Figure \ref{section2ss1fig3} depicts the absorption coefficient of the configuration C1 when the layer material is the foam S1 for $\theta^i \in[0,\pi/3]$ ($\psi^i=0$) and for $\psi^i\in[0,\pi/4]$ with $\theta^i=\pi/4$. For symmetry reason, performing calculation for larger $\psi^i$ is useless. The absorption coefficients for $\psi^i=\pi/3$ and $\psi^i=\pi/6$ when $\theta^i=\pi/4$, were found to be identical, which provides another validation of the method. The frequency of excitation of the trapped mode is slightly modified when $\theta^i$ increases: $\nu^t$ increases when $\theta^i$ increases. The absorption is nearly total up to $\theta^i\approx \pi/3$. At fixed $\theta^i$, $\psi^i$ only influences the results for frequencies higher than the first Bragg frequency. Similar results were found for the other inclusion shapes.
\section{Conclusion}\label{section4}
The influence of the periodic embedment of non-resonant three-dimensional simple shape inclusions (cube, cylinder, sphere, cone and torus) in a rigidly backed porous layer modeled in the rigid frame approximation is studied numerically and experimentally. Similarly to the 2D case, the absorption coefficient of these structures is enhanced, because of the excitation of different type of modes in particular at low frequency because of the excitation of a trapped mode that traps the energy between the inclusions and the rigid backing. This entrapment is nearly total for a specific filling fraction which is larger in 3-D than in 2-D and the frequency of excitation of this trapped mode goes down when the filling fraction increases. The FE results are validated experimentally in case of cubic inclusions. Focusing on the absorption enhancement at low frequency, a cube is better than a sphere in a cubic unit cell because it allows a larger filling fraction. It is shown that in some cases, it is impossible to obtain a nearly total absorption peak by embedding spheres because the required filling fraction cannot be reached. At a fixed filling fraction and position of the barycenter, the absorption coefficients are identical below the first Bragg frequency for the various non-resonant inclusions. In other words, for only convex inclusions, the absorption coefficient only depends on the filling fraction and position of the barycenter and not on the shape of the inclusions below the first Bragg frequency. The torus required a lower filling fraction in order to reach a nearly total absorption peak at a frequency which is smaller than for the other shapes. Differences in terms of absorption coefficient are particularly noticeable at higher frequencies than the first Bragg frequency, and allow to classify the inclusion shape, either possessing faces parallel to the interface porous/air or the the rigid backing, or not. In particular, horizontal cylinders and torus lead to larger absorption coefficients than the other shape inclusions. The so-designed structures are obviously anisotropic. The trapped mode is poorly affected by the angle of incidence, when compared to the modified mode of the backed layer. At a fixed elevation angle of incidence, the trapped mode is not affected by the azimuthal angle. These results offers large persepctives in terms of absorption enhancement of porous layer through optimization procedures and embeddement of 3D resonant inclusions.
\appendix
\section{Numerical model}\label{section2ss3}
\subsection{Finite element formulation}\label{sec:fem}

The weak form associated to the Helmholtz equation \eqref{eq:Helmholtz}, required for FE resolution, is
\begin{multline}\label{eq:weakH}
-\int_{\Omega^{\alpha}} \frac{1}{\rho^{\alpha}} \nabla \bar{q}^{\alpha} \cdot  \nabla p^{\alpha} \, \mathrm{d} \Omega  +
\int_{\Omega^{\alpha}} \frac{(k^{\alpha})^2}{\rho^{\alpha}} \bar{q}^{\alpha} p^{\alpha} \, \mathrm{d} \Omega = \\ -\int_{\partial \Omega^{\alpha}} \frac{1}{\rho^{\alpha}}\bar{q}^{\alpha} \nabla p^{\alpha}\cdot \mathbf{n} \, \mathrm{d} \Gamma
\end{multline}
for all the test functions $q^\alpha$, $\alpha=a,p$. The bar $\bar{q}^{\alpha}$ denotes the complex conjugate of $q^{\alpha}$. 

The solution being periodic, the pressure fields $p^{\alpha}$ and the test function $q^{\alpha}$ are demodulated so as to use the periodic part of the pressure denoted $\hat{p}^{\alpha}$ such that $\hat{p}^{\alpha} =\mathrm{e}^{-\mathrm{i} \mathbf{k}^{i}_\bot \cdot \mathbf{d} } p^{\alpha}$, and of the test function $\bar{\hat{q}}^{\alpha} = \mathrm{e}^{\mathrm{i} \mathbf{k}^{i}_\bot \cdot \mathbf{d} } \bar{q}^{\alpha}$. Introducing these expressions in the weak form \eqref{eq:weakH} leads to 
\begin{multline}\label{eq:weakHPeriodic}
-\int_{\Omega^{\alpha}} \frac{1}{\rho^{\alpha}} \bar{\hat{\nabla}} \bar{\hat{q}}^{\alpha} \cdot  \hat{\nabla} \hat{p}^{\alpha} \, \mathrm{d} \Omega  +
\int_{\Omega^{\alpha}} \frac{(k^{\alpha})^2}{\rho^{\alpha}} \bar{\hat{q}}^{\alpha} \hat{p}^{\alpha} \, \mathrm{d} \Omega = \\ -\int_{\partial \Omega^{\alpha}} \frac{1}{\rho^{\alpha}}\bar{\hat{q}}^{\alpha} \hat{\nabla} p^{\alpha} \cdot \mathbf{n} \, \mathrm{d} \Gamma,
\end{multline}
where the shifted gradient operator reads as $\hat{\nabla}= \nabla -\mathrm{i} \mathbf{k}^i_\bot$, $\bar{\hat{\nabla}}=\nabla +\mathrm{i} \mathbf{k}^i_\bot$.

The normal derivative of the pressure vanishes on the bottom surface $\Gamma_0$ and on $\Gamma_{inc}$ because the scatterer is infinitely rigid. The boundary term pairs on the lateral boundaries $\Gamma_r$, $\Gamma_l$ and $\Gamma_b$, $\Gamma_f$ vanish due to the periodicity of $\hat{p}$ and $\hat{q}$. The pressure and the normal velocity are continuous on the coupling boundary between the air and the porous material $\Gamma_L$,
\begin{align}
p^{a} &= p^{p} , \\
\frac{1}{\rho^{a}} \frac{\partial p^{a}}{\partial n} &=
\frac{1}{\rho^{p}} \frac{\partial p^{p}}{\partial n} .
\end{align}
The second condition is automatically accounted for by removing the boundary integral on $\Gamma_L$. The continuity of the pressure or more precisely of $\hat{p}^{a} = \hat{p}^{p}$ is ensured by the use of Lagrange multiplier $\lambda$ and its associated test function $\varsigma$. To do that, the two following integrals are evaluated on the boundary $\Gamma_L$
\begin{equation}
\int_{\Gamma_L} \bar{\varsigma} ( \hat{p}^{a} -\hat{p}^{p}) \, \mathrm{d} \Gamma + \int_{\Gamma_L} ( \bar{\hat{q}}^{a} -\bar{\hat{q}}^{p}) \lambda  \, \mathrm{d} \Gamma .
\end{equation}
Note that using Lagrange multiplier is not mandatory here and algebraical condition or penalization can be used. The main interests are to deal with Hermitian matrix and to ease the implementation.

For the sake of computation, the radiating boundary $\Gamma_\infty$ of height  $L_{\infty}$ that truncate $\Omega^{a}$ is introduced, Fig. \ref{section2ss1fig1}. The radiation of the elementary cell can be handled in the FE method with (i) DtN map, (ii) PML or (iii) modal expansion. Note PML is not suitable for the long wave limit ($\lambda \gg d $) and are not safe for this application. In this paper, the last solution is preferred for its robustness and because the modal coefficient are required to compute the absorption coefficient. In practice, only a few modes are propagative. The FE degree of freedom on $\Gamma_\infty$ are removed from FEM matrix in favor of modal amplitude. The boundary term on $\Gamma_\infty$ is easily computed thanks to Floquet mode orthogonality. On the plane boundary $\Gamma_\infty$, the total pressure reads $p^{a}=p^i+p^\infty$ and the scattered pressure can be expanded as
\begin{equation}
	\label{eq:PFloquet}
		p^\infty(\mathbf{x},\omega)\vert_{\Gamma_\infty} =  \sum_{m,n\in\mathbb{Z}^2} A_{mn} \phi_{mn},
\end{equation}
with 
\begin{equation}
\phi_{m,n}= \frac{1}{\sqrt{S}}\mathrm{e}^{\mathrm{i}\left(k_{1m} x_1 + k_{2n} x_2 \right)},
\end{equation}
where $A_{mn}$ are the amplitudes of the Floquet mode $(n,m)$, $k_{1m} = k^i_1 + m\frac{2\pi}{d}$, $k_{2n} = k^i_2 + n\frac{2\pi}{d}$, $k_{3mn}^{a} = \sqrt{(k^{a})^2 -k_{1m}^2 - k_{2n}^2}$, and $S=d^2$ is the surface of the elementary cell. To satisfy the radiation condition, i.e. the field remains bounded when $x_3 \rightarrow \infty$, the values of $k_{3mn}^{a}$ are chosen to consider both propagative and evanescent waves in $\Omega^{a}$.

It is worth noting that the FE discretization of $\Omega^{a}$ encapsuled by $\Gamma_\infty$ is not mandatory if the interface $\Gamma_L$ is a plane surface. In this case, the radiation condition can be applied directly on $\Gamma_L$ instead of $\Gamma_\infty$. The general formulation proposed here can tackle with corrugated porous material surface.

The modal expansion for the periodic part of the pressure field reads as
\begin{equation}\label{eq:pinfty}
\hat{p}\vert_{\Gamma_\infty} = \sum_{m,n\in\mathbb{Z}^2} A_{mn} \hat{\phi}_{mn} + A^i \mathrm{e}^{\mathrm{i} k_{3}^{ai} L_{\infty} },
\end{equation}
with
\begin{equation}
\hat{\phi}_{mn}= \frac{1}{\sqrt{S}}\mathrm{e}^{\mathrm{i}\left(m\frac{2\pi}{d_1} x_1 + n\frac{2\pi}{d_2} x_2 \right)}.
\end{equation}
The modal profile is changed but the value of $k_{3mn}^{a}$ remains the same.
The coefficient $A_{mn}$ can be used to compute the absorption of the material, see sec.~\ref{sec:abs}. These coefficient are cast in the vector $\mathbf{A}$.

The weak formulation arising from \eqref{eq:weakHPeriodic} yields after FE discretization (the boundary integral on $\Gamma_\infty$ will be stated later)
\begin{equation}
	\label{eq:KU}
	\mathbf{V}^{^{\mathrm{t}}} \mathbf{K}  \mathbf{U}  = \mathbf{V} \mathbf{F} .
\end{equation}
The unknowns vector $\mathbf{U}$ can be cast into a vector $\mathbf{U}_\infty$ containing the FE degree of freedom (dof) on the radiation boundary $\Gamma_\infty$ and a vector $\mathbf{U}'$ containing the other dof. The unknown vector can be expressed with \eqref{eq:pinfty}
\begin{equation}
\label{eq:U}
	\begin{pmatrix}
	  \mathbf{U}'\\
	  \mathbf{U}_\infty
	\end{pmatrix} = 
\underbrace{	\begin{pmatrix}
	\, \mathbf{I}  & \, \mathbf{0}          \\
	\, \mathbf{0} &  \mathbf{P}^\infty \\
	\end{pmatrix}}_{\mathbf{T}}
	\begin{pmatrix}
	  \mathbf{U}'\\
	  \boldsymbol{\mathbf{A}}\\
	\end{pmatrix} 
	+	
\underbrace{	\begin{pmatrix}
	\, \mathbf{0}  & \, \mathbf{0}         \\
	\, \mathbf{0} &  \mathbf{P}^i  \\
	\end{pmatrix}}_{\mathbf{G}}
	\begin{pmatrix}
	 \, \mathbf{0} \\
	  A^i
	\end{pmatrix} .
\end{equation}
Here, $\mathbf{P}$ stands for the modal projection matrix. The Floquet modes are stored in columns and the raw contains the nodal value. The same form is chosen for the test function
\begin{equation}
\label{eq:Uetoile}
	\mathbf{V} = \mathbf{T}^*
	\begin{pmatrix} 
	\mathbf{V}'\\
	\boldsymbol{B}
	\end{pmatrix},
\end{equation}
where $\mathbf{V}'$ and $\mathbf{B}$ are associated to the FE and to the modal dof, respectively.

Introducing \eqref{eq:U} and \eqref{eq:Uetoile} in \eqref{eq:KU} leads to the modified system
\begin{equation}
	\label{eq:NewKU}
\underbrace{	\mathbf{T}^\dag \mathbf{K} \mathbf{T} }_{\tilde{\mathbf{K}}}
	\underbrace{\begin{pmatrix}
	  \mathbf{U}'\\
	  \boldsymbol{A}
	\end{pmatrix}}_{\tilde{\mathbf{U}}}  = 
	\underbrace{	\mathbf{T}^\dag \mathbf{F} - 	\mathbf{T}^\dag \mathbf{K} \mathbf{G} 
		\begin{pmatrix}
	  \, \mathbf{0}\\
	  A^i
	\end{pmatrix}}_{\tilde{\mathbf{F}}} ,
\end{equation}
where $\dag$ is for the hermitian transpose. The boundary term on $\Gamma_\infty$, using the modal expansion from \eqref{eq:pinfty} and for the associated test function from \eqref{eq:Uetoile}, yields
\begin{multline}
\int_{\Gamma_\infty} \frac{1}{\rho^{a}}\bar{q}^{a} \nabla p^{a}\cdot \mathbf{n} \, \mathrm{d} \Gamma = \\
\sum_{m,n\in \mathbb{Z}^2} \bar{B}_{mn}\frac{\mathrm{i} k_{3mn}^{a}}{\rho^{a}}  A_{mn} -  \bar{B}_{00} \frac{\mathrm{i} k^{ai}_3 \sqrt{S} e^{-\mathrm{i} k_{3}^{ai} L_{\infty} }}{\rho^{a}}A^i.
\end{multline}
These terms can be easly added at the end of $\tilde{\mathbf{F}}$ and on the diagonal of $\tilde{\mathbf{K}}$. The last step is to solve the modified FE matrix $\tilde{\mathbf{K}} \tilde{\mathbf{U}} = \tilde{\mathbf{F}}$ with a sparse solver\cite{Mumps}.

\subsection{Absorption computation}\label{sec:abs}
Once the wave amplitudes have been evaluated, the integration of the acoustic intensity leading to the energy balance can be done. This integration is performed over the unit cell using the orthogonality relation of the Floquet modes. In practice, the number of propagating modes in $\Omega^{a}$ is very small and often reduced to the fundamental mode $(m,\, n) =(0,\, 0)$ (specular reflection) and the first modes. 

Thanks to the conservation of the energy, the absorbed power is given by $\mathcal{P}_{abs} = \mathcal{P}_i - \mathcal{P}_r$, where the reflected power in the $x_3$ direction is
\begin{equation}
	\label{eq:Pr}
	\mathcal{P}_r = \sum_{m,n\in\mathbb{Z}^2}\textrm{Re}\left( k_{3mn}^{a}\right) \|A_{mn}\|^2/(\rho^{a} \omega),
\end{equation}
to the incident power is
\begin{equation}
	\label{eq:Pi}
	\mathcal{P}_i =S \|A^i\|^2 k^{ai}_3  /(\rho^{a} \omega).
\end{equation}
The absorption coefficient is then defined as the ratio of the absorbed power to the incident power
\begin{equation}
	\label{eq:alpha}
	\mathcal{A}=\frac{\mathcal{P}_{abs}}{\mathcal{P}_i}=\frac{\mathcal{P}_i - \mathcal{P}_r}{\mathcal{P}_i}.
\end{equation}

\subsection{Implementation note}
The implementation of the proposed method has been performed with the open source softwares. The  finite element library \texttt{FreeFEM++}  \cite{FreeFEM} is used (version 3.20) with linear (P1) or quadratic (P2) lagrangian tetrahedral finite element, periodic boundary conditions and parallel computing facilities. The meshes were realized with \texttt{Gmsh} \cite{Geuzaine:2009} (version 2.7) with coincident mesh constraint on each opposite latteral sides of the elementary cell. The inclusions have been designed with \verb FreeCAD  \cite{FreeCAD}  (version 0.13).

\end{document}